\documentclass[12pt,preprint]{aastex}
\usepackage[yyyymmdd,hhmmss]{datetime}
\usepackage[margin=1.0in]{geometry}
\usepackage{amsmath}
\usepackage{gensymb}
\usepackage{natbib}

\shorttitle{Runaway RSG in M31}

\slugcomment{Accepted by the AJ Sep 22, 2015}

\shortauthors{Evans \& Massey}

\begin{document}

\title{A Runaway Red Supergiant in M31\altaffilmark{*}}

\author{Kate Anne Evans\altaffilmark{1, 2} \and Philip Massey\altaffilmark{1, 3}}

\altaffiltext{*}{Observations reported here were obtained at the MMT Observatory, a joint facility of the University of Arizona and the Smithsonian Institution. This paper uses data products produced by the OIR Telescope Data Center, supported by the Smithsonian Astrophysical Observatory.}
\altaffiltext{1}{Lowell Observatory, 1400 W Mars Hill Road, Flagstaff, AZ 86001; kevans@caltech.edu; phil.massey@lowell.edu.}
\altaffiltext{2}{Research Experience for Undergraduate participant during the summer of 2015.  Current address: California Institute of Technology, 1200 East California Blvd, Pasadena, CA 91125; kevans@caltech.edu. }
\altaffiltext{3}{Department of Physics and Astronomy, Northern Arizona University, Flagstaff, AZ, 86011-6010.}
\begin{abstract}

A significant percentage of OB stars are runaways, so we should expect a similar percentage of their evolved descendants to also be runaways.  However, recognizing such stars presents its own set of challenges, as these older, more evolved stars will have drifted further from their birthplace, and thus their velocities might not be obviously peculiar.  Several Galactic red supergiants (RSGs) have been described as likely runaways, based upon the existence of bow shocks, including Betelgeuse.  Here we announce the discovery of a runaway RSG in M31, based upon a 300~km~s$^{-1}$ discrepancy with M31's kinematics.  The star is found about 21\arcmin (4.6~kpc) from the plane of the disk, but this separation is consistent with its velocity and likely age ($\sim$10~Myr).  The star, J004330.06+405258.4, is an M2~I, with $M_V=-5.7$, $\log L/L_\odot$=4.76, an effective temperature of 3700~K, and an inferred mass of 12-15$M_\odot$.  The star may be a high-mass analog of the hypervelocity stars, given that its peculiar space velocity is probably 400-450~km~s$^{-1}$, comparable to the escape speed from M31's disk.

%\vskip -10pt
\end{abstract}


\vskip 30pt

\section{Introduction}
\label{Sec-intro}

\vskip -10pt
Massive stars form in giant molecular clouds, creating OB associations \cite[e.g.,][]{lada}, the members of which share similar space velocities.   OB stars that are spatially close to one of these associations, but with discrepant radial velocities, were first noted by \citet{blaauw}, who termed these  ``runaways" and proposed a possible explanation for their origin:  if the primary in a binary system sheds a significant fraction of its mass (say, in a SN explosion), then the secondary would be set loose with nearly its orbital velocity. This would cause the star's radial velocity to disagree with that of other members of the association, and over time, would cause the star to move away from its fellows.  A radial velocity study of a large sample of Galactic runaway OB stars by \citet{Gies} effectively ruled out this explanation, and instead favored a dynamical evolution explanation, a scenario that has been recently supported by the simulations of \citet{fujii} and \citet{oh}. 

Regardless of their origins, a significant percentage of OB stars are considered runaways \citep[10-50\% according to][and references therein]{Gies}.   Yet, few {\it evolved} massive stars have ever been identified as runaways, presumably as these stars are older, and as a star moves farther away, it begins to lose the context of its birth association. Without that context, unusual velocities become harder to distinguish.  Nevertheless, a few red supergiants (RSGs)  have been identified as runaways due to the presence of  bow shocks \citep{nora, mackey}. Bow shocks are known to be present for some OB runaways \citep{norb}, and so have been considered evidence that a star is a runaway.   Galactic RSGs considered to be runaways include Betelgeuse, $\mu$ Cep, and IRC-10414 \citep{gvar}.

While conducting a radial velocity study of RSGs in M31, we discovered that the RSG J004330.06+405258.4 is not only quite isolated, but also possesses a radial velocity 300~km~s$^{-1}$ at odds with that expected from its location.  We conclude that this is a runaway RSG, the first identified in another galaxy, and the fastest known.

\section{Observations}
\label{Sec-obs}

\begin{figure}
\epsscale{0.8}
\plotone{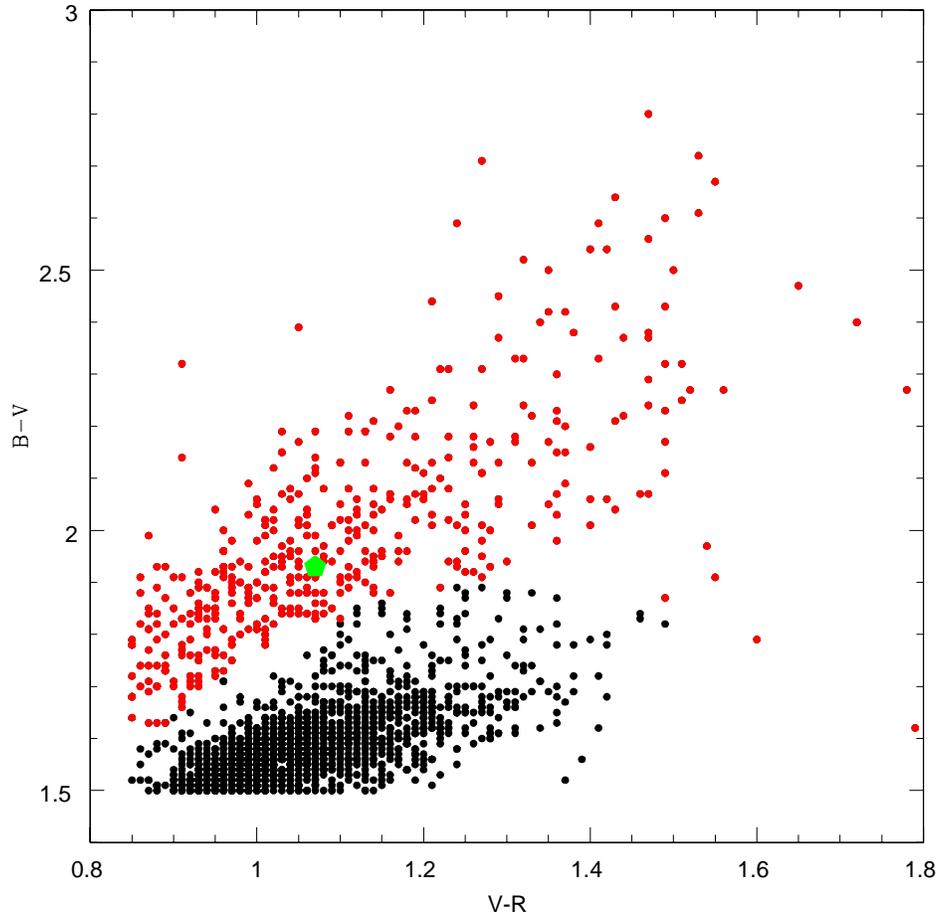}
\epsscale{1.0}
\caption{\label{fig:PhotSplit} Two-color diagram of red stars observed with Hectospec. The objects in red are suspected red supergiants, and the objects in black are suspected foreground stars. The assumed division between the two is given by  Equation~\ref{eq:Para}. J004330.06+405258.6 is represented by the green pentagon.}
\end{figure}

J004330.06+405258.4 was one of many M31 RSGs we observed spectroscopically.  Our sample came from \citet{massey09}, who had identified candidate RSGs and foreground stars in M31 using the photometry of the Local Group Galaxy Survey \citep[LGGS,][]{massey06}.  For very red stars, {\it V-R}  remains a temperature discernment, but {\it B-V} becomes primarily an indicator of surface gravity due to the effects of line-blanketing in the {\it B} bandpass \citep{massey06}.  Thus, red supergiants can be separated from foreground stars in a {\it B-V}, {\it V-R} diagram.  Candidate RSGs were separated from the candidate foreground stars using the following equation:
\begin{equation}
\label{eq:Para}
B-V=-1.599(V-R)^2+4.18(V-R)-0.83
\end{equation}
We summarized the available photometry for J004330.06+405258.4 in Table~\ref{tab:Phot}.  The location of J004330.06+405258.4 in the two-color diagram of our sample is shown in Figure~\ref{fig:PhotSplit}, where we have colored the expected RSGs in red and the expected foreground stars in black, based upon the above cut-off. The location of our runaway star, J004330.06+405258.4 is very much in the expected RSG area.  

The full results of the radial velocity study will be discussed elsewhere, but here we will summarize the data relevant to J004330.06+405258.4. The object was observed using Hectospec \citep{hecto}, a 300 optical fiber-fed spectrograph on the 6.5-m MMT telescope. We used the 270 line mm$^{-1}$ grating, which is blazed at 5000~\AA, and covers a spectral range from 3650 to 9200~\AA.  The grating provides a dispersion of 1.2 \AA\, pixel$^{-1}$ and a spectral resolution of 6.2 \AA.  Some of the fibers were assigned to blank sky to be used for sky subtraction, and calibration included HeNeAr and quartz lamp exposures. Our runaway was observed during four different nights in the Fall of 2014 in queue mode for 90 minutes of exposure each, as summarized in Table~\ref{tab:All}. Following the observations, data were passed through the SAO pipeline, and the wavelength zero-points  adjusted slightly by using night sky lines.  The wavelengths were then corrected to a heliocentric reference frame.

As no blocking filter could be used, we expect contamination by second-order blue light at the longer wavelengths of our spectrum.  However, at the Ca\, {\sc ii}  triplet lines (8498, 8542, 8662 \AA), which we use for radial velocities, we expect this contamination will be only $\sim$3\%, as  $B-R\sim2.9$ (Table~\ref{tab:Phot}), implying that the overlapping second-order blue at 4250~\AA\ will be roughly a factor of 15 smaller in flux at 8500~\AA,  and the light dispersed by an additional factor of 2 in second order.  

Flux calibration was provided by observations of the spectrophotometric standard Feige 34 midway through the semester. The resulting sensitivity curves were kindly provided by Nelson Caldwell.  Our experience is that this is typically good to 5-10\% in the fluxes relative to wavelengths (i.e., the colors), although occasionally much larger errors may be present \citep[see also][]{hecto2}.

\section{Analysis}
\label{Sec-Analysis}

\vskip -8pt

The first step in our larger project is to establish whether the stars selected from the photometry by \cite{massey09} were truly RSGs in M31 or merely foreground stars.  Except for stars in the NE section of the galaxy, the combination of the $\sim$250~km~s$^{-1}$ rotational velocity and $\sim$$-300$~km~s$^{-1}$ systemic velocity \citep{deV} makes it easy to separate foreground Milky Way dwarfs from {\it bona fide} M31 members using radial velocities,  as shown by \citet{drout} for yellow supergiants. 

The observed radial velocities were calculated using {\sc xcsao}, a cross correlation tool, in IRAF\footnote{IRAF is distributed by the National Optical Astronomy Observatory, which is operated by the Association of Universities for Research in Astronomy (AURA) under a cooperative agreement with the National Science Foundation.}. For our velocity templates, we used 21 Hectospec spectra of six M31 RSGs for which radial velocities were already known \citep{massey09}, and we restricted the cross correlation to the Ca\, {\sc ii} triplet lines, which are very strong in RSGs; this avoids using the extremely broad molecular features. We then averaged the result for each spectra, weighting the results in accord with the the internal errors. For stars with more than one observation, the results of each spectra were averaged, with weights assigned by the weighted errors of each spectra. Full details will be given in the larger paper that is in preparation.

We then compared the observed radial velocities with that expected from the star's location in M31 using the simple kinematic model adopted by \citet{drout}, which is based upon the seminal study by \citet{rubin}.  In general, the expected radial velocity $V_{\rm exp}$ from circular rotation will be $V_{\rm exp}=V_0+V(R)\sin\xi \cos\theta,$ where $V_0$ is the systemic radial velocity, V(R) is the rotational velocity at a distance from the galactic center (R) within the plane of the disk, $\xi$ is the angle between the line of sight and the perpendicular to the plane of the galaxy, and $\theta$ is the angle from the semi-major axis.  Thus, $\cos\theta$ is $X/R,$ where $X$ is the position along the major axis.   The gross simplification that V(R) is a constant (which is equivalent to saying that dark matter dominates the kinematics) works remarkably well \citep{drout,massey09}, leaving us then with a linear relationship.  For consistency with these earlier works, and to simplify using the \citet{rubin} data, we have adopted $\xi$=77$^\circ$ and then used
$V_{\rm exp} = -295 +241.5 (X/R)$.   This equation does an excellent job of fitting both the velocities of the H\,{\sc ii} regions measured by \citet{rubin} as well as the RSG radial velocities measured by \citet{massey09}, as shown in Figure 2 of the latter work.  

\begin{figure}
\epsscale{0.8}
\plotone{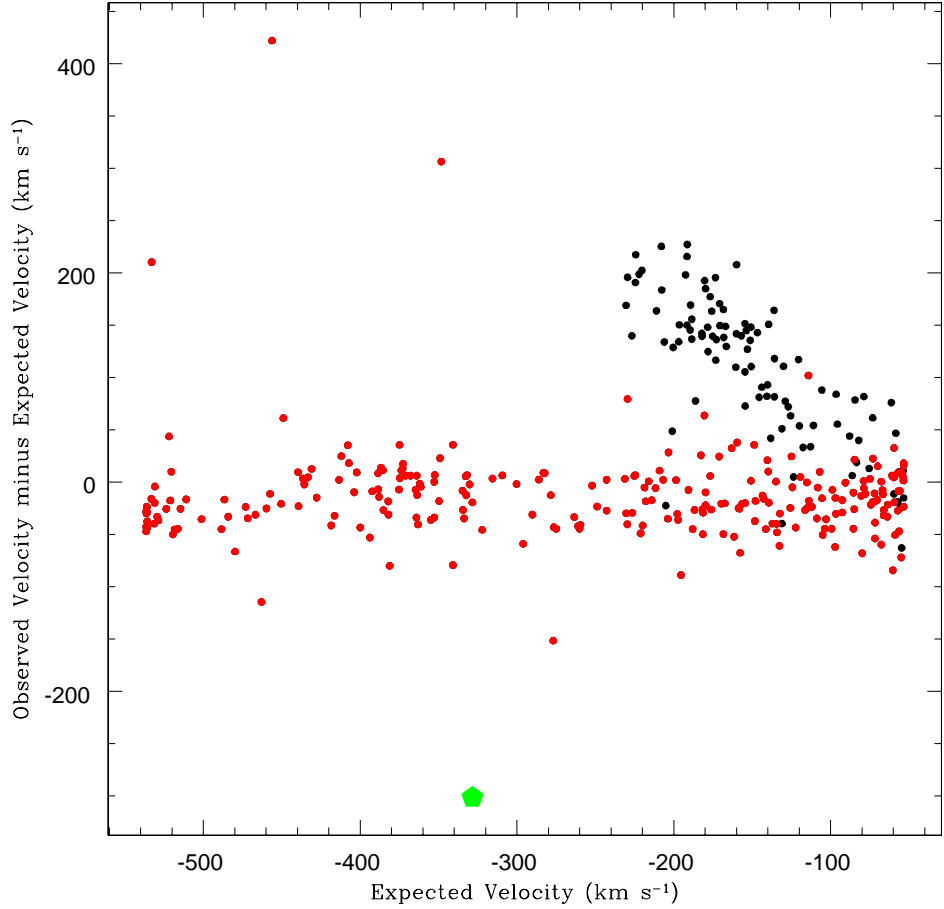}
\epsscale{1.0}
\caption{\label{fig:Major}  Difference between the observed velocity ($V_{\rm obs}$) and the expected velocity ($V_{\rm exp}$).   The stars expected to be RSGs on the basis of their location in the two-color diagram (Figure~\ref{fig:PhotSplit}) are shown with red symbols; as expected, these cluster around a velocity difference of 0.  The stars expected to be foreground stars based upon their photometry are shown as black symbols, and as as expected these cluster around around a line with slope -1 in this diagram. The runaway star is noted with a green pentagon.}
\end{figure}

We compare the velocity difference ($V_{\rm obs} - V_{\rm exp}$) for each star in Figure~\ref{fig:Major}. The suspected RSGs and the suspected foreground stars split generally well into two distinct groups. The RSGs cluster around a velocity difference of zero, indicating that their observed velocities are about equal to the velocities that we expect based on M31's kinematics. The foreground stars have velocity differences that are about equal to the negative of $V_{\rm exp}$, as their $V_{\rm obs}$ values are about zero. 
It is immediately clear that one star stands out as having a very peculiar negative velocity.  This star is J004330.06+405258.4, which we have highlighted with a green pentagon.   There
are a few other outliers: three purported RSGs that are obviously foreground stars (due to being marginal in the 2-color diagram), and several stars whose photometry was compromised by crowding and/or whose spectra were poor.  These will be discussed in more detail in our larger paper.  

The radial velocities of J004330.06+405258.4 are given in Table~\ref{tab:All}, and we can see that the four measurements are quite consistent.   Furthermore, the spectra of this star are well exposed, and the triplet lines readily measured; see Figure~\ref{fig:Spect}.  
The observed radial velocity of J004330.06+405258.4 is $-630$~km~s$^{-1}$, while its expected radial velocity is $-328$~km~s$^{-1}$, leading to a $-302$~km~s$^{-1}$ peculiar radial velocity.  

Consistent with its peculiar radial velocity, J004330.06+405258.4 is also very well separated spatially from other massive stars, as shown in Figure~\ref{fig:Image} from the LGGS.  We measure a separation of about 21$\arcmin$ (4.6~kpc) from the major axis.   Is this separation reasonable?  Assuming an age of 10~Myr,  and a tangental velocity similar to the peculiar component of its radial velocity (i.e., $-$300~km~s$^{-1}$), we can expect J004330.06+405258.4 to have traveled $\sim$3~kpc from its birthplace, an extremely good match to what we observe\footnote{It takes 11~Myr for a 15$M_\odot$ star to being core He-burning, and 15~Myr for a 12$M_\odot$ according to the Geneva evolutionary models \citep{Sylvia}}.

\subsection{Physical Properties  of the Runaway}

We assigned a spectral type of M2~I to J004330.06+405258.4.  M-type subclasses are determined primarily by the depth of the TiO bands.  Rather than normalize the spectra (a problematic issue for very red stars where there is only ``pseudo" continua), we instead compared the spectra of J004330.06+4052858.4 to stars that had been previously classified by \citet{levesque} digitally in log space.  In Figure~\ref{fig:Class} we show that the spectrum is well matched to that of M2~I star BD+59\degree38.

We can determine the effective temperature $T_{\rm eff}$ of J004330.06+405258.4 by using the MARCS stellar models \citep{marcs}, following the same procedure as in \citet{massey09}.  We adopted a $\log g$ of 0.0, and compared the spectral features (primarily the depths of the molecular bands) to those of models of different effective temperatures, adjusting the color excess as needed.  A ``good" fit was determined using the TiO bands between about 6000 and 7000\AA.  An example is shown in Figure~\ref{fig:TempFit}. The results of these fits are shown in Table~\ref{tab:All}.  The temperature of 3700 K can be compared to the 3675~K temperature found by \citet{massey09} as the median of M2~I stars in M31 using the same super-solar metallicity models. 

With this value for the effective temperature we can now determine the bolometric luminosity.    We measured $E(B-V)$ values
of 0.15-0.20 from our fitting.  This amount of reddening is consistent with the 
typical 0.13 value found from OB stars by \citet{massey07}, but given the star's location far from any star-forming region, is probably indicative of some circumstellar component, as is commonly found
with RSGs \citep{Smoke}.  Adopting $A_V=3.1 \times 0.15 = 0.46$, and a distance of 760~kpc \citep[from][]{vandenbergh2000}, we find an absolute
visual magnitude $M_V=-5.65$.  The bolometric correction at {\it V} for the 3700~K MARCS model is $-1.51$, giving us a bolometric magnitude $M_{\rm bol}=-7.16$, or $\log L/L_\odot$ of 4.76.  Alternatively, we can use the 2MASS photometry (given in Table~\ref{tab:Phot}) to find the bolometric luminosity.  We expect $A_K\sim 0.112 \times A_V$ \citep{schlegel} or  0.05.  Thus, $M_K = -9.93$.  The bolometric correction at {\it K} for a star of 3700~K is +2.76 according to the MARCS models.  Thus from the K-band, we derive $M_{\rm bol} =  -7.17$, in near perfect agreement with the value derived from $V$.
A comparison with the evolutionary tracks of \citet{Sylvia} gives an inferred mass of J004330.06+405258.4 of 12-15$M_\odot$.

\section{Discussion}
\label{Sec-sum}

\begin{figure}
\epsscale{0.8}
\plotone{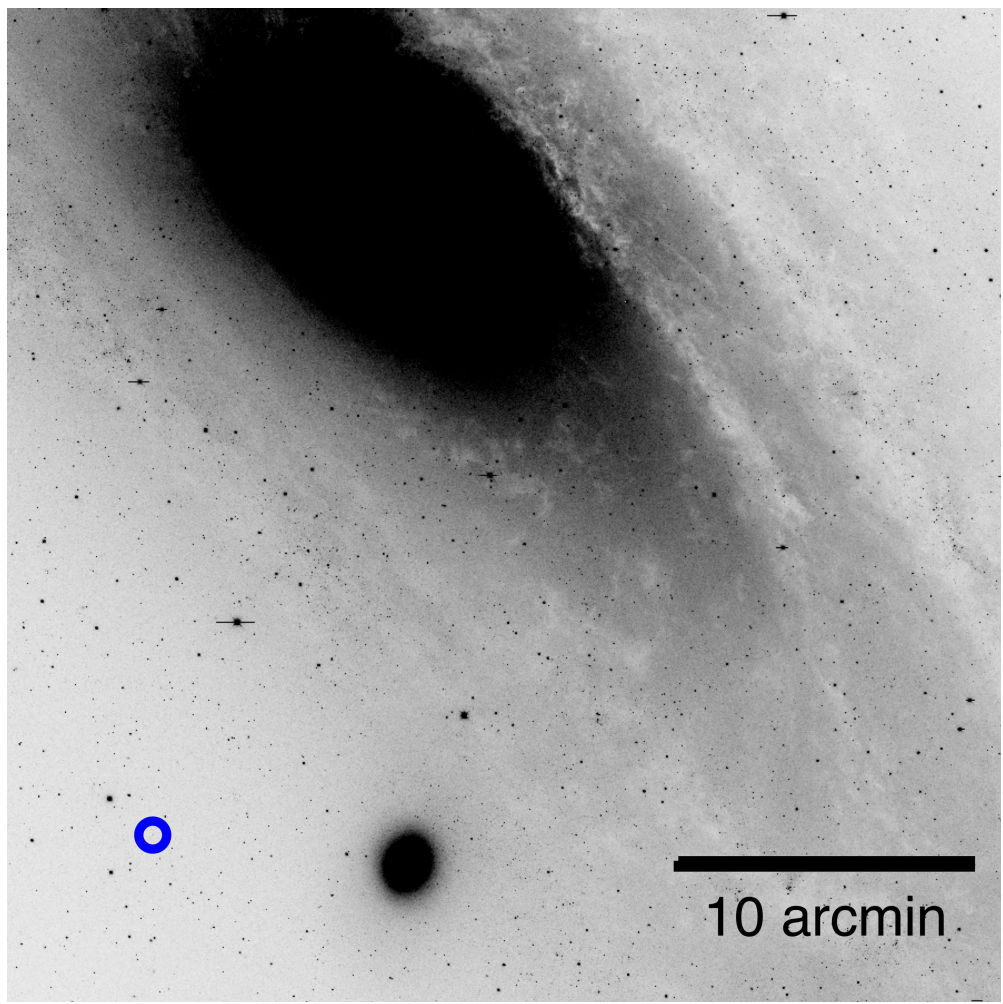}
\epsscale{1.0}
\caption{\label{fig:Image} Location of J004330.06+405258.4 in M31.  Here J004330.06+405258.4 is highlighted with a blue circle. The distance from the runaway to the semi-major axis is about 21$\arcmin$ (4.6~kpc). This image was taken in the $V$-band as part of the LGGS.  N is up and E is to the left.}
\end{figure}

\begin{figure}
\epsscale{0.8}
\plotone{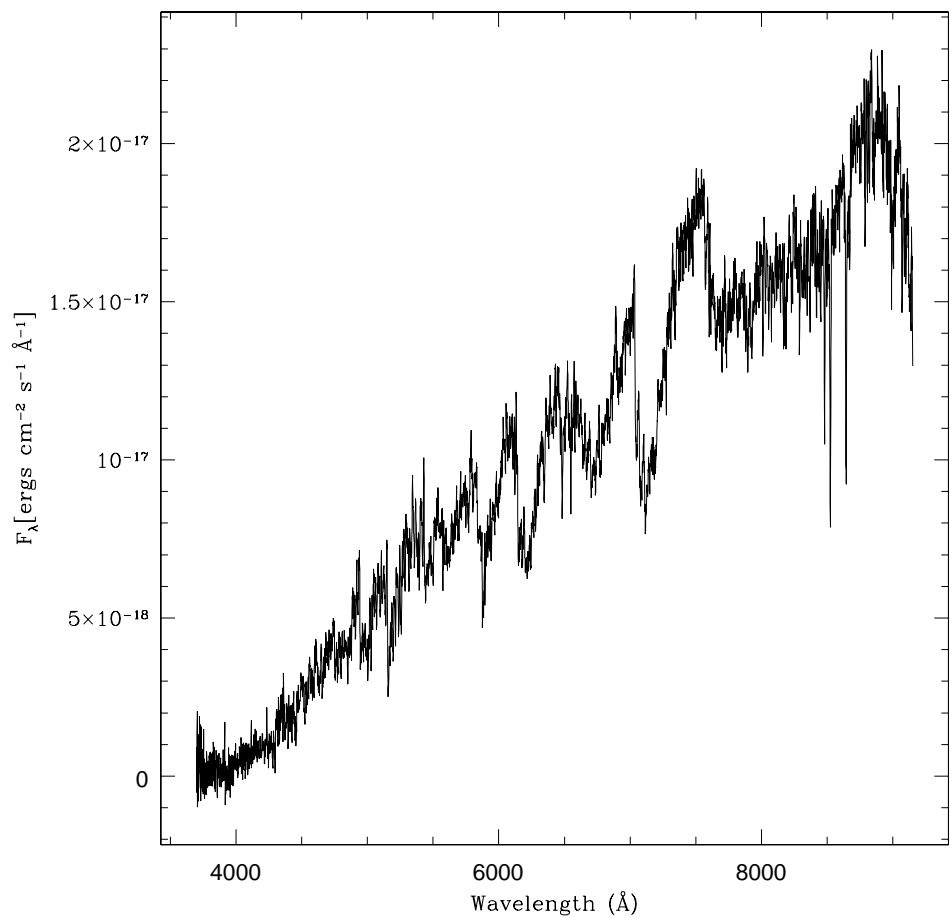}
\epsscale{1.0}
\caption{\label{fig:Spect}  Spectrum of J004330.06+405258.4.  The Ca\, {\sc ii} triplet used for the radial velocities are clearly visible at 8498-8662\AA.}
\end{figure}

\begin{figure}
\epsscale{0.8}
\plotone{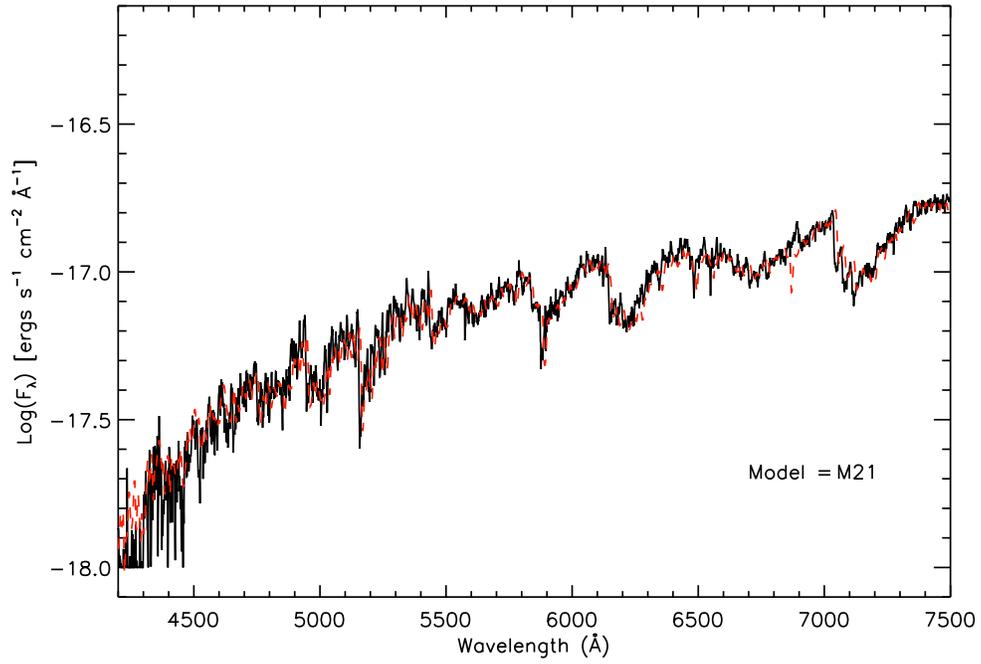}
\epsscale{1.0}
\caption{\label{fig:Class} Comparison with an M2~I Spectral Standard.  A spectrum of J004330.06+405258.4 (black) is compared with a spectrum of BD+59$^\circ$ 38 (red).  The latter was classified as M2~I by \citet{levesque}. }
\end{figure}

\begin{figure}
\epsscale{0.8}
\plotone{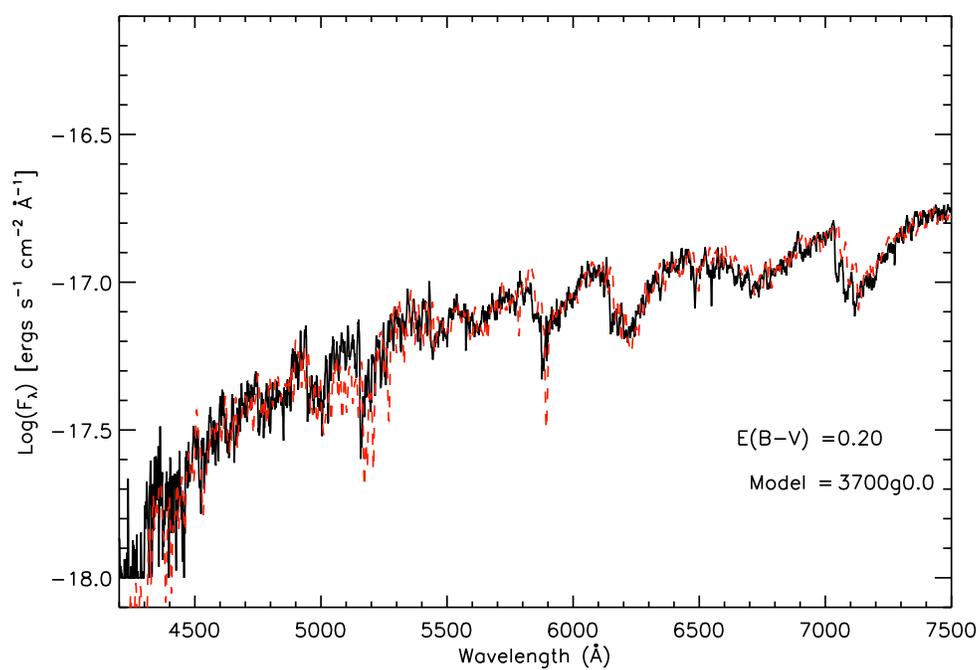}
\epsscale{1.0}
\caption{\label{fig:TempFit} Model Fit. A spectrum of J004330.06+405258.4 (black) is compared with a 3700~K, $\log$ g=0.0 MARCS model from \citet{massey09}.}
\end{figure}

We have established that J004330.06+405258.4 is runaway RSG, the first such star clearly identified in another galaxy.  Furthermore, with a peculiar radial velocity of -300~km~s$^{-1}$, it is the fastest known runaway massive star of which we are aware.   Given its distance, a direct measurement of its tangental component
via proper motions is not practical given current methods, but the spatial separation from M31's disk (Figure~\ref{fig:Image}) suggests that the tangential component of the velocity is similar, and thus that the peculiar space velocity is 400-450~km~s$^{-1}$. 

Lower mass ``hypervelocity" stars are known in our own Galaxy; these are stars that are moving at peculiar velocities of $>$500~km~s$^{-1}$, and, because of their much greater ages, have traveled many tens of kpc in their lifetimes \citep[see, e.g.,][]{hyper}.  They are escaping from the Galaxy.  Our M31 RSG runaway may be a high-mass analog of such stars, rather than related to traditional runaways, given that its velocity is so much greater than the 30~km~s$^{-1}$ that usually use to distinguish an OB runaway \citep{blaauw}.    We note that the peculiar velocities of the other known Galactic RSGs are quite modest by comparison:  IRC-10414 at 70~km~s$^{-1}$\citep{gvar}, $\mu$ Cep at 22~km~s$^{-1}$\citep{cox}, and Betelgeuse at 56~km~s$^{-1}$\citep{nora}.  Similarly only two O-type stars in the classic \citet{Cruz} catalog have peculiar radial velocities $>$100~km~s$^{-1}$, and both are
under 120~km~s$^{-1}$.

We were curious what evidence there was of similar objects in other galaxies.  In their study of yellow supergiants (YSGs) in the SMC, \citet{NeugentSMC} do, in fact, note that one star, J01020100-7122208, has 
an anomalously high radial velocity, +307~km~s$^{-1}$, compared to the SMCÕs systemic velocity of 158~km~s$^{-1}$.  They suggest the star is either a binary or a runaway; based on a single observation, they cannot tell, but it is clear from their Figure~7 that the starÕs radial velocity stands out by 100~km~s$^{-1}$ from the other YSGs.  Similarly, the histograms of the radial velocities of yellow and red supergiants in the LMC studied by  \citet{NeugentLMC} also show two interesting outliers (their Figure 4): there is a YSG, J04530398-6937285, that has a radial velocity of +373~km~s$^{-1}$, and a RSG, J04482407-7104012 (CPD-71\degree285), with a radial velocity of +401~km~s$^{-1}$, both of which can be compared to the LMCÕs systemic velocity of +278~km~s$^{-1}$.  Again, further observations would be needed to establish that these velocities are not high due to binary motion.  Amongst the RSGs in M33, there is an intriguing example, J013403.34+302611.7, with a radial velocity that is discrepant by 160~km~s$^{-1}$ \citep[see Figure 8 in][]{DroutM33}. None of these are, of course, as extreme as the case of our M31 RSG runaway. 

Could J004330.06+405258.4 be a lower mass object?  Any alternative explanation needs to account not only for the starÕs radial velocity but also its photometry.   We can essentially rule out the star being a foreground dwarf:  an M2 Galactic dwarf at $V=19.2$ would have to be at 0.7~kpc to be this faint, and with a radial velocity of $-630$~km~s$^{-1}$ it would have to be a hypervelocity star; an unlikely coincidence given how close it is in the plane of the sky to M31's disk.  (We would also not expect it to be as reddened.)   What if it were an asymptotic giant branch (AGB) star in M31's halo?  The bolometric magnitude $M_{\rm bol}=-7.2$ ($\log L/L_\odot = 4.76$) is faint enough that we expect some contamination by intermediate-mass AGBs in this magnitude range.  However, the typical velocity dispersion of halo stars in our own Milky Way is about 100~km~s$^{-1}$ \citep{BrownMW}, and we expect that M31's is not much different.  We would then still be left trying to explain why the star had a radial velocity $330$~km~s$^{-1}$ more negative than M31's $-300$~km~s$^{-1}$ systemic velocity, as this is far in excess than what we expect for a typical halo object.   

J004330.06+405258.4 is seen in relative close proximity (in projection) to the galaxy M32, the compact low-luminosity elliptical visible $\sim$8\arcmin\  to the west in Figure~\ref{fig:Image}.  Could there be a connection?  It is well known that dwarf ellipticals have very little or no current star formation.  In the case of M32, about 3\% of its population consists of 
``young" (ages $<$2~Gyr), metal-rich stars \citep{TodM32}, but there is no evidence of any massive stars. Nor is it likely that J004330.06+405258.4 is an  AGB associated with M32, as the galaxy's radial velocity is -205 km s$^{-1}$ \citep{deV}, which would leave us an even greater discrepancy in radial velocity.

Is J004330.06+405258.4 escaping from M31?  This question is largely moot, as the star will only live another million years or so before undergoing a SN explosion; in that time it will only move only another 400-450 pc. 
We do note, however, that it is moving sufficiently fast to have escaped the gravitational attraction of the disk.   Although the gravitational well of M31 is complex, with a massive dark-matter halo, we can make a crude estimate simply using the mass of the disk. \citet{rubin} calculates that the mass within 24 kpc of the center of M31 is $\sim 2 \times10^{11}$ M$_{\odot}$.    The projected distance of J004330.06+405258.4 is 5.6 kpc.  We would thus expect naively the escape velocity is {\it on the order} of 560~km~s$^{-1}$.  (We note as a reality check that the escape velocity for the sun from the Milky Way is 550~km~s$^{-1}$ according to \citealt{RAVE}.)  Thus it is not unreasonable that J004330.06+405258.4 has gotten to where it is today.

The presence of bow shocks in association with runaway Galactic RSGs  \citep{mackey} leaves a question as to whether or not we could detect such a structure around our M31 runaway. According to \cite{nora}, Betelgeuse has a bow shock 0.8 pc in size. At the distance of M31, a similiar-sized bow shock would extend about  0.2\arcsec.  This size is large enough to be readily detectable, at least using space-based imaging.  However, it remains to be seen what the effect of our runaway's increased speed and the likely decreased hydrogen column density would be on bow shock size and presence.

\acknowledgements
  
We are grateful to the Steward Observatory Time Allocation Committee  for their generous allocation of observing time on the MMT,  and to Perry Berlind, Mike Calkins, and Marc Lacasse for their excellent support of Hectospec.  Nelson Caldwell kindly provided the calibration files for flux calibrating our data, as well as managing the difficult task of queue scheduling our program.  Kathryn Neugent helped take some of these observations as part of another project, and also provided very useful scientific suggestions on this project. Emily Levesque and Joe Llama provided useful help with some of the coding for fitting the MARCS models. An anonymous referee kindly made useful suggestions which improved this paper. This publication makes use of data products from the Two Micron All Sky Survey, which is a joint project of the University of Massachusetts and the Infrared Processing and Analysis Center/California Institute of Technology, funded by the National Aeronautics and Space Administration and the National Science Foundation (NSF).  K. A. E.'s work was supported through the NSF's Research Experiences for Undergraduates program through Northern Arizona University and Lowell Observatory (AST-1461200), and P. M.'s were partially supported by the NSF through AST-1008020 and through Lowell Observatory. A grant from the Mt.\ Cuba Astronomical Foundation for computing facilities is gratefully acknowledged.

{\it Facilities:} \facility{MMT(Hectospec)}

%\clearpage

\bibliographystyle{apj}
\bibliography{masterbib}

\clearpage

\begin{deluxetable}{l r c c l l }
%\tabletypesize{\scriptsize}
\tablecaption{\label{tab:Phot} Photometry of J004330.06+405258.4}
\tablewidth{0pt}
\tablehead{
\colhead{Measurement}
& \colhead{Value} 
& \colhead{Error} \\
& \colhead{(mags)} & \colhead{(mags)} \\
}
\startdata                        
{\it V}\tablenotemark{a} & 19.212 & 0.005 \\
{\it B-V} \tablenotemark{a} & 1.933 & 0.013 \\
{\it V-R}\tablenotemark{a} & 1.077 & 0.006 \\  
{\it K}\tablenotemark{b} & 14.525 & 0.087 \\
{\it J-K}\tablenotemark{b} & 1.076 & 0.106 \\
\enddata
\tablenotetext{a}{From the LGGS \citep{massey06}.}
\tablenotetext{b}{From 2MASS \citep{2MASS}}.
\end{deluxetable}

\begin{deluxetable}{l c c c c c}
%\tabletypesize{\scriptsize}
\tablecaption{\label{tab:All} Radial Velocities and Results of the Model Fitting}
\tablewidth{0pt}
\tablehead{
\colhead{HJD}
& \multicolumn{2}{c}{Radial Velocity (km s$^{-1}$)} 
& 
& \multicolumn{2}{c}{MARCS fitting}  \\ \cline{2-3} \cline {5-6}
& \colhead{$V_{\rm obs}$}
& \colhead{$\sigma$}
&
& \colhead{$T_{\rm eff}$}
& \colhead{$E(B-V)$} 
}
\startdata                        
2456926.254 & -625.9  & 1.02  && 3725 K & 0.15 \\
2456982.310 & -633.7  & 1.12  && 3700 K & 0.20 \\
2456987.260 & -623.6  & 1.73  && 3700 K & 0.15 \\  
2456989.177 & -632.7  & 1.01  && 3700 K & 0.20 \\  
Adopted Values & -630 & 1 && 3700 K & 0.15 \\
\enddata
\end{deluxetable}

\clearpage

\end{document}